\documentclass[a4paper, amsfonts, amssymb, amsmath, reprint, showkeys, nofootinbib, twoside]{revtex4-1}
\usepackage[english]{babel}
\usepackage[utf8]{inputenc}
\usepackage[colorinlistoftodos, color=green!40, prependcaption]{todonotes}
\usepackage{amsthm}
\usepackage{mathtools}
\usepackage{physics}
\usepackage{xcolor}
\usepackage{graphicx}
\usepackage[left=23mm,right=13mm,top=35mm,columnsep=15pt]{geometry} 
\usepackage{adjustbox}
\usepackage{placeins}
\usepackage[T1]{fontenc}
\usepackage{lipsum}
\usepackage{csquotes}
\usepackage[pdftex, pdftitle={Article}, pdfauthor={Author}]{hyperref} 
\begin{document}
\title{Rethinking scale in network neuroscience:\\Contributions and opportunities at the nanoscale}

\author{Richard F. Betzel$^{1,2}$}
\email[Correspondence email address: ]{rbetzel@umn.edu}
\author{Caio Seguin$^{3}$}
\author{Maria Grazia Puxeddu$^{4}$}
\affiliation{1 - Department of Neuroscience, University of Minnesota, Minneapolis MN 55419}
\affiliation{2 - Masonic Institute for the Developing Brain, University of Minnesota, Minneapolis MN 55413}
\affiliation{3 - Department of Psychiatry, University of Melbourne, Melbourne AUS}
\affiliation{4 - Department of Computer, Control, and Management Engineering, University of Rome La Sapienza, Rome, Italy}

\date{\today} 

\begin{abstract}

Network science has been applied widely to study brain network organization, especially at the meso-scale, where nodes represent brain areas and edges reflect interareal connectivity inferred from imaging or tract-tracing data. While this approach has yielded important insights into large-scale brain network architecture, its foundational assumptions often misalign with the biological realities of neural systems. In this review, we argue that network science finds its most direct and mechanistically grounded application in nanoscale connectomics—wiring diagrams reconstructed at the level of individual neurons and synapses, often from high-resolution electron microscopy volumes. At this finer scale, core network concepts such as paths, motifs, communities, and centrality acquire concrete biological interpretations. Unlike meso-scale models, nanoscale connectomes are typically derived from individual animals, preserve synaptic resolution, and are richly annotated with cell types, neurotransmitter identities, and morphological detail. These properties enable biologically grounded, mechanistically interpretable analyses of circuit structure and function. We review how nanoscale data support new forms of network modeling, from realistic dynamical simulations to topology-informed circuit inference, and outline emerging directions in multimodal integration, cross-species comparisons, and generative modeling. We also emphasize the continued importance of meso- and macro-scale connectomics, especially in human neuroscience, and discuss how nanoscale insights can inform interpretation at coarser scales. Together, these efforts point toward a multi-scale future for network neuroscience, grounded in the strengths of each resolution.

\end{abstract}


\maketitle

\section*{Introduction} \label{sec:introduction}

\begin{figure*}[t]
	\centering
    \includegraphics[width=1\textwidth]{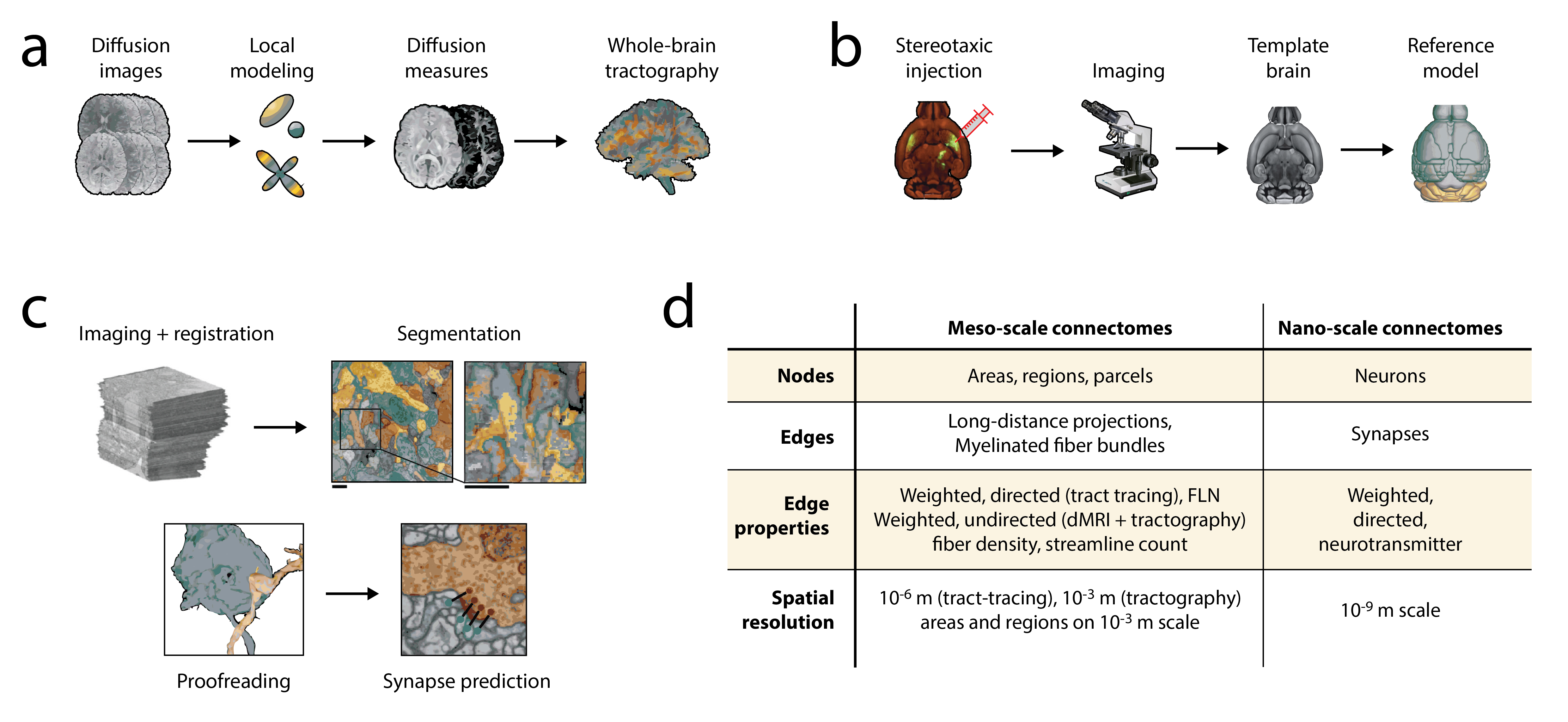}
	\caption{\textbf{Connectome processing and reconstruction pipelines.} High level overview of pipelines for constructing connectomes at different resolutions using different modalities. Schematic pipelines for connectome construction based on (\emph{a}) diffusion MRI pipeline, (\emph{b}) tract tracing, and (\emph{c}) electron microscopy. Panel \emph{a} adapted from \cite{beaudoin2021modern}. Panel \emph{b} adapted from \cite{oh2014mesoscale}, and panel \emph{c} adapted from \cite{dorkenwald2022flywire}.}  \label{fig0}
\end{figure*}

Network science provides a powerful abstraction for representing complex systems, modeling their behavior, and uncovering their organizing principles \cite{barabasi2016network, newman2018networks}. In neuroscience, networks have helped map the architecture of nervous systems across many scales—from synaptic microcircuits \cite{varshney2011structural, white1986structure} to systems-level interactions among cortical and subcortical areas \cite{hagmann2008mapping, power2011functional}. But not all scales are equally well suited to network modeling. When the fundamental assumptions of network science—nodes and edges as discrete, meaningful entities; connectivity as directed, weighted, and grounded in biological transmission—are violated or obscured, the interpretability and explanatory power of the model diminishes.

Nanoscale connectomes, which represent the connectivity of individual neurons through synapses reconstructed from high-resolution electron microscopy with nano-meter resolution \cite{winding2023connectome, dorkenwald2024neuronal}, fulfill the assumptions of network models to a far greater degree than meso-scale \cite{oh2014mesoscale, markov2014weighted, bota2015architecture} or macro-scale connectomes \cite{hagmann2008mapping}. At this resolution, edges correspond to anatomically and functionally meaningful units: synapses with known polarity, weight, and sometimes neurotransmitter identity. Nodes are neurons whose types, morphologies, and spatial locations are often known. These features enable network analyses to be biologically grounded, interpretable, and mechanistically testable \cite{barabasi2023neuroscience}.

In this review, we focus primarily on nanoscale connectomes that are both complete, synapse-resolved, and structurally reconstructed—criteria that currently restrict our attention to a small number of model organisms, notably \emph{Caenorhabditis elegans} and \emph{Drosophila melanogaster}. These datasets are unique in providing fully mapped circuits at single-synapse resolution across the entire brain or nervous system of an individual animal. While efforts to perform similar reconstructions in other species are ongoing -- see, for example, partial reconstructions in mouse \cite{microns2025functional} and human cortex \cite{shapson2024petavoxel} and in larval zebrafish \cite{hildebrand2017whole, du2025central, li2025multiplexed, petkova2025connectomic} -- these remain far from the completeness required to achieve the same level of interpretability.


Network science first gained widespread adoption in the analysis of meso- and macro-scale connectomes, where tools such as diffusion MRI enabled the construction of structural brain networks in living humans. These studies revealed hallmark features like small-world topology, modular organization, and inter-individual variation in hub connectivity—all of which proved informative for understanding development, disease, and cognition \cite{bullmore2009complex, sporns2013network}. However, such networks are built on indirect measures and group-level averaging, which can obscure fine-grained mechanisms of circuit computation \cite{papo2014complex, papo2025biological, ross2024causation, bertolero2020nature}. As connectomics enters the nanoscale era, network science has the opportunity to operate at a resolution that aligns more tightly with its assumptions—and with the biology it seeks to explain.

Here, we discuss how fundamental properties of complex systems--such as paths and modules--manifest at this scale, and how the rich metadata accompanying nanoscale connectomes is inspiring innovative network analyses. Our goal is to illustrate how network approaches at the nanoscale can reveal mechanistic insights into neural computation, thus bridging the gap between abstract network models and the biological circuits they represent.

\section*{Main} \label{sec:results}

\begin{figure*}[t]
	\centering
    \includegraphics[width=1.0\textwidth]{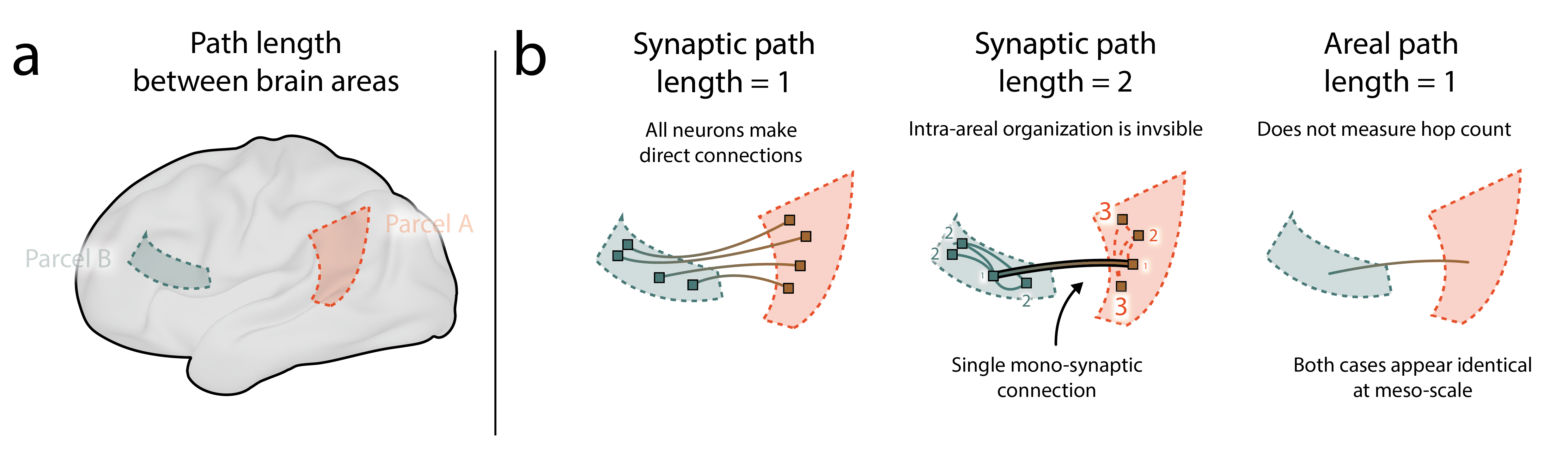}
	\caption{\textbf{Interpretational issues with inter-areal hops and paths are resolved at nano-scale.} The compatibility of path-based measures at the nano-scale is clearest when we view its incompatibility with meso-scale connectome data. Consider the problem of estimating the number of hops between two areas/parcels, A and B. If information about individual neurons and their synaptic connectivity was available, we could resolve all the neuron-to-neuron with perfect fidelity -- i.e. we could count the number of neurons on the shortest path from pre-synaptic neurons in parcel A to any post-synaptic neuron in parcel B. Panels \emph{b},left and \emph{b},middle show two configurations where the length of shortest paths from parcel A to B differ. However, these configurations are not distinguishable if the only information that is available is areal connections (see panel \emph{b},right).}  \label{fig1}
\end{figure*}

\subsection*{Data Fidelity and Ground Truth}

Mesoscale networks, such as those derived from diffusion MRI, offer invaluable access to the human and mammalian brains. Human mesoscale networks are acquired non-invasively and are reproducible and scalable, allowing for large-cohort studies that link network properties to cognition, development, and disease \cite{van2013wu, casey2018adolescent}. But these networks are necessarily statistical summaries, inferred through fiber tracking models and heuristics \cite{jeurissen2019diffusion, pestilli2014evaluation}. They average across time, individuals, and measurement noise. Nodes may not correspond to functionally coherent units \cite{gordon2016generation}, and edges reflect estimated, rather than observed, interactions. Further, because these networks are acquired non-invasively and reconstructed inferentially, there is no ground truth against which these estimates can be compared and scored.

In non-human animals, mesoscale networks can be reconstructed from invasive procedures and connections defined based on observation. However, these networks represent composites, by construction, and are assembled from many different individuals and experiments stitched together to approximately tile an entire brain, mapping connections at millimeter scale \cite{oh2014mesoscale, markov2014weighted}.

By contrast, nanoscale connectomes offer direct, high-fidelity, neuron-resolution measurements of connectivity (Fig.~\ref{fig0}). Reconstructions based on electron microscopy (EM) or high-resolution volume imaging provide synapse-level data, often across entire circuits \cite{scheffer2020connectome} or brains \cite{dorkenwald2024neuronal, winding2023connectome}. Each edge in these networks reflects a countable number of chemical synapses between identified neurons. This enables rigorous validation of network models, including those pertaining to flow, communication, centrality, and motifs. For instance, in the \emph{Drosophila} hemibrain, all connections are directional and weighted by synapse count, allowing for accurate estimation of effective paths and bottlenecks \cite{scheffer2020connectome} \footnote{While nanoscale connectomes present reconstructions at neuron/synapse level, offering unique opportunities to connect with biology, like meso-scale connectomes, errors in the reconstruction pipeline propagate to biases in the connectome. For instance, automated image segmentation, labeling, and annotations are error-prone (can mistakenly merge or split neurons) and require time-consuming manual proofreading \cite{dorkenwald2022flywire}. Similarly, sampling biases may lead to connectomes dominated by ``easy-to-reconstruct'' neurons, failing to accurately reconstruct neurons exhibiting long-range, complex aborization.}.

The realism of nanoscale connectomes makes them uniquely valuable for network science. Because the data are derived from individual animals, they preserve idiosyncratic features such as rare cell types \cite{schlegel2024whole}, local motifs \cite{lin2024network}, or asymmetries across hemispheres—features that would be lost in population-averaged mesoscale datasets\footnote{At present, fully reconstructed nanoscale connectomes exist for only a small number of individuals in each species. It remains unclear to what extent these datasets are representative of the species as a whole, and caution is warranted when generalizing findings from these connectomes to broader principles of brain organization.}. This fidelity is especially important for modeling neural computation, where function may hinge on the presence or absence of specific microcircuits \cite{lappalainen2024connectome, caravelli2023mean, di2016nano}. Moreover, because they are directly observed, nanoscale networks offer a fertile test bed for validating network metrics against experimental perturbations or dynamic recordings.

Another critical advantage of nanoscale connectomics is the ability to study how synaptic connectivity interfaces with the chemical and molecular environment of the nervous system. Brains are not composed of neurons and synapses alone; neuromodulators, neuropeptides, and extrasynaptic signals shape circuit function in powerful ways. At the mesoscale, such influences are largely invisible. But in nanoscale connectomes, particularly those of \emph{C. elegans}, researchers have begun to characterize the molecular signatures of neurons and the spatial organization of non-synaptic contacts that support chemical communication. For example, recent work has mapped the neuropeptidergic signaling architecture of the \emph{C. elegans} nervous system, revealing a layered, multi-channel form of communication that complements the structural synaptic graph \cite{bentley2016multilayer, ripoll2023neuropeptidergic}. Likewise, extrasynaptic and volume-transmission signaling mechanisms, including those mediated by dense-core vesicles, have been systematically characterized using EM and molecular annotation datasets \cite{taylor2021molecular, randi2023neural}. These findings emphasize that nanoscale connectomes are uniquely suited to studying how synaptic wiring operates within—and is modulated by—the broader neurochemical context of the organism.

Importantly, the meso-to-nano fidelity gap does not diminish the value of mesoscale data. Rather, nanoscale connectomes complement them, offering a high-fidelity neural substrate for interpreting the patterns observed at coarser resolutions. As models of brain function strive to bridge levels of organization, having reliable data at the microscale is essential for building and validating multiscale models \cite{betzel2017multi}.

\begin{figure*}[t]
	\centering
    \includegraphics[width=1.0\textwidth]{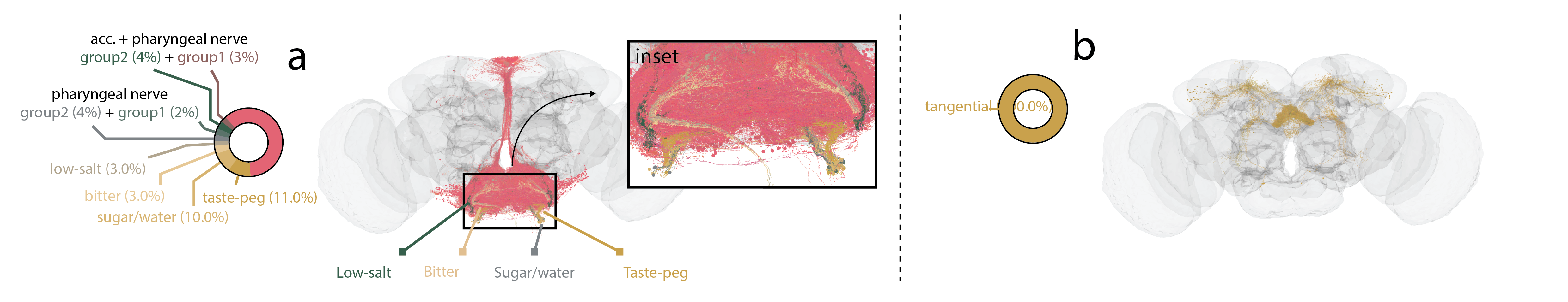}
	\caption{\textbf{Community detection for circuit (re)discovery.} Community detection algorithms identify groups or sub-graphs of neural elements. Generally, the elements assigned to a given sub-graph exhibit some shared pattern of connectivity; i.e. there exists some statistical regularity in terms of how the interact with one another and the rest of the connectome. We envision community detection being a useful tool for nanoscale connectomics. First, it can help reduce the dimensionality of nanoscale connectomes, shifting focus away from massive numbers of neurons to far fewer communities while still preserving the overall network structure. Second, communities can be further characterized by comparing them against annotations or experimental data. Sometimes these comparisons are well-aligned. In panel \emph{b}, we show a community comprised of neurons with the same annotational label (\texttt{sub\textunderscore class} assignment of ``tangential cell''). In this case, the community is well-aligned with expert-curated labels of these neurons. In panel \emph{a}, we show a community composed of neurons whose labels include multiple different annotation types (though all broadly related to gustation). In this case, the community is not as neatly aligned with the underlying annotations and demands further characterization.}  \label{fig2}
\end{figure*}

\subsection*{Paths, Walks, and Communication}

One of the most powerful tools in network science is the ability to quantify how information flows through a system. In many domains, this is formalized through paths—sequences of edges that connect nodes through a network. But the interpretability of such paths depends critically on what edges represent. In meso- and macro-scale brain networks, edges often reflect correlations or tractography estimates between regions, without a clear model to link them to underlying axonal or synaptic connections. That is, in the absence of a ground truth against which tractography can be compared and due to the inferential nature of tractography algorithms, it is unclear whether a direct tractographic connection is equivalent to a direct synaptic connection (the same reasoning holds for poly-synaptic pathways). This distinction is critical if one adheres to the hypothesis that computation is carried about by circuits of neurons as opposed to areas of spatially co-localized neurons. As a result, shortest paths or random walks in meso-scale networks may be mathematically interesting but can be biologically ambiguous. Moreover, given the inexact correspondence with synaptic connectivity and the coarseness of the parcels/areas by which labeled neurons \cite{markov2014weighted, oh2014mesoscale} or streamlines \cite{bijsterbosch2020challenges} are aggregated, it is unclear what a step or hop in a path even represents. In addition, it is possible that neuronal pathways with different lengths (number of synaptic hopes) can give rise to identical interareal distances (see Fig.~\ref{fig1}b).

Nanoscale networks, in contrast, offer a literal interpretation of paths (Fig.~\ref{fig1}). Each edge is a synapse, and a path represents a potential polysynaptic route for signal transmission \cite{dvali2024diverging}. This grounds communication models anatomically, allowing researchers to simulate or infer actual flows of neural activity. For example, studies in C. elegans have shown that path-based models can predict behavioral outputs and identify critical interneurons \cite{varshney2011structural}. In Drosophila, paths from sensory to descending neurons reveal how distinct pathways converge to control motor behavior \cite{winding2023connectome, schlegel2021information}. These analyses gain traction because the constituent edges represent real, quantifiable synapses, often annotated by polarity and neurotransmitter type, allowing for models that are not merely formal but mechanistic.

This interpretability also extends to models of control \cite{yan2017network}, flow \cite{seguin2023brain, van2012high}, and influence \cite{van2011rich}. Centrality measures such as betweenness, closeness, or eigenvector centrality become biologically testable: neurons with high centrality may serve as hubs, integrators, or bottlenecks in circuits of neurons, as opposed to the coarser interareal variety. In nanoscale networks, such claims can be evaluated against recordings, perturbations, or behavioral consequences, enabling a closer integration of theory and experiment. This is not to say that measures of distance and communicability (a weighted sum of all multi-step walks \cite{estrada2008communicability, crofts2009weighted}) are not practically useful when measured at the meso-scale; rather, their biological interpretation is less straightforward, complicated by the incompleteness of tract-tracing (with no guarantee that the majority of projections have been labeled) and the known reconstruction limits of tractography.

Indeed, a major strength of nanoscale connectomes is their potential to bridge theoretical and experimental neuroscience. Because they are derived from model organisms amenable to genetic and optical manipulation, they open the door to empirical tests of network-level hypotheses. For example, in \emph{C. elegans}, connectome-derived predictions about circuit dynamics can be tested \emph{via} optogenetic stimulation of individual neurons and high-speed behavioral tracking \cite{nguyen2016whole, randi2023neural}. Similarly, studies in \emph{Drosophila} have used targeted silencing or ablation of neurons and connections to assess the role of specific pathways in behavior \cite{imambocus2022neuropeptidergic}. These interventions are often interpreted using network concepts such as control centrality or minimal driver sets \cite{yan2017network}, allowing for a uniquely mechanistic loop between model, perturbation, and behavior. Such experimental tractability remains unattainable in human connectomics, making nanoscale systems uniquely powerful for causal network neuroscience.

\subsection*{Communities, Modules, and Functional Subcircuits}

Community detection algorithms—used to uncover modular structure in networks—are a cornerstone of network science \cite{fortunato2010community, newman2012communities}. Identifying modules in mesoscale brain networks has not only advanced our understanding of large-scale brain organization \cite{meunier2009hierarchical, meunier2010modular, bassett2010efficient, sporns2016modular, puxeddu2025leveraging} but also provides valuable targets for biomarker discovery. Modules often align with known functional systems \cite{power2011functional, yeo2011organization}, such as sensory, motor, or default-mode networks, while connector hubs enable integration across modules \cite{hagmann2008mapping}, supporting higher-order cognitive functions and behavior \cite{suarez2020linking}. Alterations in modular organization have been linked to development \cite{gu2015emergence}, aging \cite{betzel2014changes}, and neurological or psychiatric disorders \cite{lynch2024frontostriatal}, including Alzheimer’s disease and autism, highlighting its potential as a biomarker for brain health and disease. These findings have advanced our understanding of large-scale brain organization. However, the resolution of mesoscale data limits the ability to identify modules at the level of specific computations or behaviors. Note that this does not preclude the possibility that communities detected at the meso-scale are relevant for biomarker generation \cite{lynch2024frontostriatal} or personalized clinical interventions \cite{nahaspersonalized}. Rather, the limitation is in terms of the scale of inquiry; dissection of the internal functions/computations of communities defined and characterized at the meso-scale is generally impossible, at least from the wiring diagram alone (experiments, however, are useful for at least constraining what these internal computations might be).

At the nanoscale, on the other hand, community detection can reveal fine-grained subcircuits whose structure is closely tied to function \cite{betzel2024hierarchical, kunin2023hierarchical} (Fig.~\ref{fig2}). For instance, in the Drosophila central brain, modular decomposition of the connectome has identified nested substructures aligned with compartments, pathways, and functional motifs \cite{schlegel2023whole}. These communities often reflect developmentally or anatomically defined regions, such as glomeruli, columns, or layers. Moreover, because the networks are directed and weighted, algorithms can distinguish between feedforward, recurrent, and feedback modules, extending existing classifications based on community interactions \cite{betzel2018diversity}.

Crucially, the discovery of communities at the nanoscale is not merely confirmatory—it can be hypothesis-generating. Network clustering can suggest candidate circuits for specific behaviors, which can then be tested \emph{via} genetic or physiological experiments. In this way, nanoscale connectomics enables a more iterative dialogue between network analysis and biological experimentation \cite{pospisil2024fly}.

\subsection*{Cell Types, Compartments, and Multiscale Annotation}

\begin{figure*}[t]
	\centering
    \includegraphics[width=1.0\textwidth]{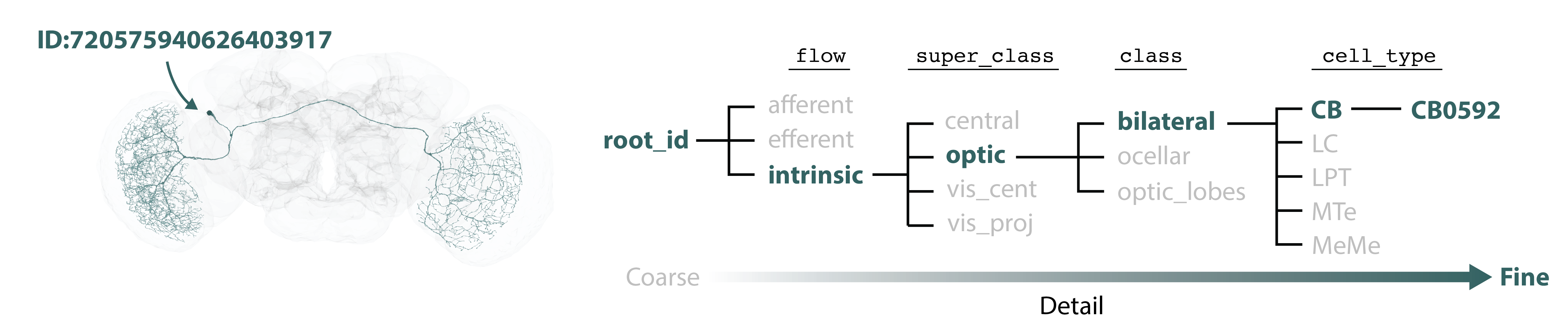}
	\caption{\textbf{Annotations promote vertex heterogeneity and extend links to biology.} Though recent efforts at the meso-scale have similarly attempted to imbue brain regions/areas with meta-data and annotations \cite{bazinet2023towards}, brain network annotations are best-deployed at the nanoscale, where they are applied to well-defined neural elements -- e.g. neurons and synapses. Here, we illustrate the depth of annotations from the FlyWire \emph{Drosophila} connectome. Neurons are associated with root IDs. They are then annotated with progressively more detail, ranging coarse descriptions of information flow (\texttt{flow}) to the cell types (\texttt{cell\textunderscore type}).}  \label{fig3}
\end{figure*}

A defining feature of nanoscale connectomics is the richness of biological annotations. Nanoscale nodes are individual neurons, often annotated with type, lineage, morphology, polarity, neurotransmitter, and compartmentalization\footnote{This type of annotation is becoming more common in mesoscale connectome analysis (see, for instance, \citet{bazinet2023towards}), though presently annotations are still coarse-grained to roughly millimeter scale and often represent composites of many brains \cite{hansen2021mapping} (or in special cases, composites constructed from very \emph{few} \cite{hawrylycz2012anatomically} or even single brains \cite{amunts2013bigbrain})} (Fig.~\ref{fig3}). This wealth of metadata enables multiscale, multimodal network analyses grounded in biology.

In the mammalian cortex, for instance, the ability to assign synapses to apical or basal dendrites of pyramidal cells has enabled detailed modeling of integrative computation across cortical layers \cite{kubota2015functional}. In insect brains, axo-dendritic polarity and neuropil compartmentalization offer a scaffold for modeling compartment-specific dynamics, such as local \emph{versus} global inhibition \cite{scheffer2020connectome}. These annotations provide a bridge between structural topology and dynamic computation, allowing network models to incorporate known rules of biophysics and physiology \cite{shiu2024drosophila}.

These annotations also allow for integrative models that combine structure, cell type, and function. For example, graph neural networks can incorporate morphological or physiological features as node attributes, enabling supervised learning of connectivity rules or functional outcomes. In turn, cell-type-specific connectivity patterns can reveal organizing principles of circuit architecture, such as convergence, divergence, symmetry, or hierarchical nesting \cite{betzel2024hierarchical}.

Furthermore, these annotations enable biologically constrained generative models \cite{newman2016structure, betzel2017generative}. By learning from annotated networks, one can build models that simulate realistic neural circuits, explore evolutionary design principles, or test the consequences of structural perturbations \cite{liu2024generative, goulas2019spatiotemporal}. Such models are increasingly important as connectomics shifts from description to prediction \cite{haspel2023reverse, rabinowitch2020would}.


\subsection*{Discussion}

Although the first whole-brain neuronal network was reconstructed decades ago \cite{white1986structure}, the recent widespread emergence of nanoscale connectomes represents a pivotal moment for network neuroscience. At this scale, the correspondence between network representations and biological mechanisms is direct: nodes are neurons, edges are countable synapses, and paths trace putative poly-synaptic routes of information flow \cite{betzel2024parallel}. As such, nanoscale networks offer an opportunity to ground theoretical constructs in biological precision, enabling models that are not only descriptive but mechanistic.

The rich annotations available in nanoscale connectomes—cell types, synapse properties, spatial compartments—enable models that go beyond topology, linking structure to computation. These data help support biologically realistic simulations, inform theories of circuit function, and constrain generative models that aim to capture the principles of neural architecture. They also open doors for cross-species comparisons, evolutionary analyses, and the construction of biologically grounded machine learning models.

The advent of the nanoscale connectomics era opens new opportunities for causal testing at the network level not possible with existing network datasets. With complete synaptic maps, researchers can predict driver sets and control motifs \cite{liu2011controllability}, then test them using optogenetic activation, targeted ablation, or chemogenetic inhibition in model organisms \cite{rocchi2022increased, lake2020simultaneous}. These closed-loop experiments—where network analysis directly motivates—and is validated by—behavioral perturbation hold promise for bridging connectivity and computation in a rigorous, empirically grounded framework.

These developments also point toward a future in which connectomes are not just anatomical maps but multimodal atlases incorporating molecular, neurochemical, and transcriptional information \cite{bazinet2023towards}. Such approaches could illuminate how synaptic motifs are embedded within neuropeptidergic or extrasynaptic signaling fields \cite{ceballos2024mapping}, enabling more realistic models of neuromodulation \cite{betzel2024controlling, kim2025inferring}, behavioral state control, and network flexibility \cite{bassett2011dynamic}. 

While we have focused on a narrow class of complete, synapse-resolved structural connectomes, this definition is not meant to exclude related efforts that trade off completeness for scale, or structure for function \cite{helmstaedter2013connectomic, shapson2024petavoxel}. Emerging datasets, such as the 1 mm$^3$ mouse visual cortex reconstruction \cite{microns2025functional}, larval zebrafish brain EM \cite{hildebrand2017whole}, whole-brain calcium imaging in transparent model organisms \cite{chen2018brain}, and neuronal cultures with synthetically-generated topologies \cite{hoffmann2024pixels, montala2022rich}, offer powerful insights despite not meeting all criteria. These intermediate-scale resources point toward a future in which nanoscale connectomics may generalize to larger brains and more complex behaviors.

In this review we have touted nanoscale connectomes as transformative for the field of network neuroscience. At the same time, it is important not to diminish the value of mesoscale and macroscale connectomics. These approaches remain indispensable for studying the human brain, especially in health and disease \cite{lynch2024frontostriatal}. Non-invasive imaging allows for longitudinal, developmental, and population-level studies that are not feasible with invasive connectomics \cite{casey2018adolescent}. Moreover, precision mapping is improving the resolution and individual specificity of mesoscale data \cite{gratton2018functional, gordon2017precision}, narrowing the gap between imaging-based networks and the underlying biology. Ultimately, understanding brain function will require integration across scales—from the molecular to the systems level—and both mesoscale and nanoscale data are essential components of that effort \cite{rosas2025top}.

While nanoscale datasets are primarily available in model organisms, they can nonetheless help shape the future of human network neuroscience. Detailed maps of cellular and microcircuit organization in animals serve as ground truth for validating tract-tracing, diffusion modeling, and network estimation techniques in humans \cite{mansour2025duke}. They also provide inspiration for generative models that bridge spatial scales—e.g., by linking macroscale connectivity patterns to stereotyped microcircuit motifs \cite{goulas2019spatiotemporal}. As human data becomes more individualized and fine-grained, insights from nanoscale connectomes can guide both interpretation and experimental design, especially in clinical settings where mechanistic understanding remains limited.

One of the most significant advantages of meso-scale connectomes is their compatibility with non-invasive methods. This allows researchers to longitudinally track changes in whole-organ, brain network architecture associated with learning, aging, and disease progression, providing a window into dynamic processes at the systems level \cite{casey2018adolescent}. Moreover, emerging techniques in precision neuroimaging are increasingly enabling subject-specific network reconstructions, offering a path toward individualized models of brain organization and pathology \cite{gordon2017precision, laumann2015functional}.

Additionally, meso-scale networks have catalyzed many foundational discoveries in network neuroscience. Concepts such as rich-club organization \cite{van2011rich}, hub vulnerability \cite{crossley2014hubs}, and the correspondence between structure and function \cite{honey2007network} were initially discovered through studies of interareal brain networks. These findings continue to inform clinical neuroscience, particularly in disorders where systems-level disruptions are evident, such as schizophrenia, Alzheimer's disease, and traumatic brain injury \cite{bullmore2009complex, fornito2015connectomics}.


A central advantage of network science is its ability to describe neural systems across scales using a common mathematical language. Graph-theoretic measures—such as centrality, clustering, communicability, and modularity—are abstract by design, making them applicable to connectomes that differ dramatically in physical size, biological substrate, or acquisition modality. This level of abstraction is rare in computational neuroscience, where models are often tightly coupled to the specific details of a given system’s dynamics, morphology, or biophysics.

Because the same algorithmic tools and measures can be used on both human and \emph{C. elegans} connectomes, network neuroscience enables direct comparisons across species and brain sizes. For example, studies have applied diffusion-based and shortest-path communication models to both human functional networks and the full \emph{C. elegans} synaptic connectome to estimate likely routes of signal flow and behavioral control \cite{stiso2018spatial, yan2017network, avena2018communication}. If these models prove predictive in both contexts, it suggests that certain principles of communication—such as redundancy, routing efficiency, or controllability—may be conserved across evolution and scale. This flexibility also allows researchers to ask whether the same network motifs or topological constraints that support cognition in humans may underlie simpler forms of behavior in smaller organisms.

Importantly, this cross-scale compatibility is not merely convenient—it offers a rare opportunity to generate and test unifying theories of brain organization. As nanoscale connectomes expand in number and complexity, they may serve as fertile testbeds for evaluating which network principles are scale-invariant, which are species-specific, and how such principles manifest in relation to behavior, development, or evolution.

In short, meso-scale connectomics remains essential for understanding brain function at the systems level. Rather than viewing meso- and nanoscale approaches as competing paradigms, they should be seen as mutually informative. Nanoscale models can provide high-fidelity, near-ground-truth biological constraints for meso-scale analyses, while meso-scale data contextualize fine-scale dynamics within the larger organizational scaffold of the brain.
\section*{Acknowledgments} \label{sec:acknowledgements}

\subsection*{Declaration of competing interests}
RFB, CS, and MGP have no conflicts.

\subsection*{Author contributions}
RFB conceived of this review, wrote the initial draft, generated figures, and wrote the final version of the manuscript as well. CS and MGP helped to edit the initial, but also wrote and edited the final version of the manuscript.

\bibliography{bibfile}

\end{document}